# Deep Learning-based ECG Classification on Raspberry PI using a TensorflowLite Model based on PTB-XL Dataset

Kushagra Sharma[1] and Rasit Eskicioglu[2]

[1]Biomedical Engineering Program, University of Manitoba, Winnipeg, Canada
[2]Department of Computer Science, University of Manitoba, Winnipeg, Canada

## ABSTRACT

*The number of IoT devices in healthcare is expected to rise sharply due to increased demand since the COVID-19 pandemic. Deep learning and IoT devices are being employed to monitor body vitals and automate anomaly detection in clinical and non-clinical settings. Most of the current technology requires the transmission of raw data to a remote server, which is not efficient for resource-constrained IoT devices and embedded systems. Additionally, it is challenging to develop a machine learning model for ECG classification due to the lack of an extensive open public database. To an extent, to overcome this challenge PTB-XL dataset has been used. In this work, we have developed machine learning models to be deployed on Raspberry Pi. We present an evaluation of our TensorFlow Model with two classification classes. We also present the evaluation of the corresponding TensorFlowLiteFlatBuffers to demonstrate their minimal run-time requirements while maintaining acceptable accuracy.*

## 1. INTRODUCTION

Over the past few years, there has been a tremendous increase in IoT devices such as smartwatches, virtual assistants, smart plugs, smart switches, and healthcare devices like intelligent insulin pumps, pulse-oximeter, and heart rate monitors. There were more than 26 billion connected devices in 2019. This number may reach 75 billion by 2025 and over 500 billion by 2030 [1] [2]. The applications of IoT devices in healthcare have also risen sharply due to significantly increased demand since the COVID-19 pandemic. The IoT-healthcare sector's compounded annual growth rate (CAGR) is above 20% for many regions. Rising awareness and AI-enabled IoT devices detecting real-time anomalies are significant growth drivers [3]. Additionally, rapidly increasing 5G network coverage will enable more connected devices as it offers a larger spectrum, better speed, and lower latency [4].

An increasingly enormous amount of data is being generated and transmitted with the addition of more devices in house and healthcare settings [5]. Since these devices are often battery operated, optimizing them for longer operational life is necessary [6]. Higher power consumption in processing or transmitting the data leads to shorter battery life and increased maintenance costs. Additionally, implantable medical devices cannot be changed that often. Therefore having a longer battery life is more crucial for implantable devices [1].

ECG signals need a high sampling rate for clinical use and for higher precision applications [7]. Offloading such data to a server for computation requires high transmission activity, negatively impacting energy efficiency. It leads to battery draining faster as the energy consumed in transmission would exceed the energy consumption for on-device processing [8]. Numerous models have been explored to mitigate these issues –IoT-fog-cloud architecture [9][10][11], Mobile Edge Computing [12], Distributed Mobile Edge Computing [10], and 'Reinforcement Qlearning Model' [13]. Therefore, computing ECG data on the IoT device using an efficient machine learning model appears to be ideal.

The heart acts as an electrical dipole. The strength and orientation of this dipole change with each beat. These changes are reflected in different ECG leads to capture unique information. The signal waveform on these leads analyzes the heart's performance and detects possible anomalies. These unique features differentiate abnormal ECG waveforms from typical ECG waveforms. Some examples are a) Ventricular Fibrillation – absence of P waves and broad QRS complex, b) Atrial Fibrillation – Normal QRS complex with immeasurable PR interval. c) First Degree AV blocks – Normal QRS with PR interval greater than 0.20 seconds [14][15].

Machine learning is employed to capture these variations in ECG signals. Numerous machine learning models were explored to classify ECG signals, and they yield great accuracy. However, they are often not suitable for applications on new data due to their limited information based on their smaller training dataset. PTB-XL dataset provides 21,837 12-lead recording from 18885 patients. PTB-XL dataset covers a wide span of signals and diagnostic classes [16]. It exposes the machine learning model to various signals that could arise in real-life application scenarios. Since the database has 12 leads, models can be developed for clinical and non-clinical applications. Clinical applications often utilize all 12 leads, whereas home users utilize 1 to 3 leads.

The literature suggests numerous ways for classifying ECG signals – CNN, RNN, SVN, K-NN, and MLP [17][18][19]. Using MATLAB simulation, AbdelhamidDaamouche et al. [20] and MiladNazarahari et al. [21] classified ECG signals using wavelets. The average accuracy of both methods is over 80%, along with other similar methods[22][23][24][25]. SerkanKiranyaz et al. [26] propose a fast and accurate method to classify ECG signals using a trained, dedicated CNN for a patient. Although they present decent accuracy, these models cannot be used for realtime applications due to their limited information about anomaly classes.

Recent literature suggests a drift towards using LSTM and CNN to classify ECG data on wearable devices [27][28]. Transfer learning is also being used to deploy pertained CNN models for on-device inference of data [29]. Wavelets are also used for feature extraction, which could be used by the LSTM model [30]. Since binary LSTMs could run on the limited memory of wearable devices [31], their application is increasingly being explored. Numerous other models are being explored and adapted for the use case scenario. This research primarily focuses on leveraging 1D-CNN for modeling using TensorFlow/Keras.

TensorFlow [32] is a free and open-source software library for machine learning. It was developed by the Google Brain team and initially released in 2015. TensorFlow utilizes tensors and graphs. A tensor is a vector or matrix representing the data - input, output, and intermediate. In simple words, a tensor is a data container. Using a tensor for operations like multiplication or addition in large data is computationally less demanding than conventional 'for' loops. In TensorFlow, connected tensors perform the computations inside a graph. It uses a graph to represent a function's computations. Using the graphs provides the advantage of running the model on multiple platforms – CPU, GPU, TPU, embedded, etc. The graphs could also be saved for later use. We have used TensorFlow's 'saved model' format in this research. TensorFlow allows converting the saved models to FlatBufferusing TensorFlowLite Converter.

TensorFlowLite [33] is an open-source deep learning framework for on-device inference [34]. With the increase in wireless technology, more applications are being explored for IoT devices requiring sophisticated deep learning models on edge and embedded devices [35][36][37]. TensorFlowLite converts the TensorFlow/Keras saved model to a FlatBuffer. A Flatbuffer is an serialization cross-platform library for C/C++/C, Java and Python. The Flatbuffer can be executed with much fewer resources. During the conversion process, quantization optimizations are used to reduce the model's size. Flatbuffers are memory efficient, cross-platform compatible, and smaller in size. They do not need to be parsed or unpacked; they are ideal for resourceconstrainedIoT devices [38].

## 2. RELATED WORK

Considerable research has already been done for processing ECG signals, laying the foundation for this work. Many machine learning models previously developed had smaller datasets that are not suitable for real-life applications. The use of machine learning models for ECG anomaly detection on IoT devices using TensorFlowLiteFlatbuffers is relatively new. The contribution of every work done previously in all the relevant domains has played a significant role in developing this work.

Survey papers by Hongzu Li et al. [39] and David Menotti et al. [40] discuss numerous methods for heart anomaly detection. Some methods rely on hardware enhancements, while others incorporate advanced software processing. Guy J. J. Warmerdam et al. [41] developed a multichannel hierarchical probabilistic framework with predictive modeling for fetal ECG R-peak detection. H. B. Seidel et al. [42] proposed using Haar-DWT hardware architecture for energyefficient processing of ECG signals while maintaining an R-peak detection. Saira Aziz et al. [43] have tested various machine learning models using the SPH database to classify various classes.

The application of wavelets has been explored extensively for signal processing. Wavelets are used for analyzing signals in the time-frequency domain while retaining the significant features of the signal. Jeong-Seon Park et al. [44] have used wavelet transform and modified Shannon energy envelope to detect R-peak. Agostino Giorgio et al. [45] discuss detecting late ventricular potentials using wavelet denoising and support vector machine classification. In this work, wavelets have been used to fix the baseline wander present in ECG signals while preserving the frequency information based on the method proposed by Arman Sargolzaei et al. [46].

The models mentioned above often need a powerful computer to run on, so they are not suitable for IoT devices. For handling the issue of computational efficiency, different offloading methods have been proposed for IoT devices. Various architectures such as distributed Mobile Edge Computing and partial offloading have been explored [12]. Distributed intelligence models [47] and IoT-fog-cloud architecture were also proposed to optimize offloading efficiency. However, these architectures are often not efficient when a relatively large amount of data is transmitted [8].

Therefore, applications are being developed efficiently using deep learning models on edge nodes and embedded devices [48][49][50][51][37]. Amit Ghosh et al.[52], developed an assistive technology for the visually impaired that can detect objects in real-time on Raspberry Pi. The wearable goggle they developed had earphones for providing feedback to the user. Reenu Mohandas et al. [53] deployed a deep learning model on Raspberry Pi for detecting face masks

International Journal of Artificial Intelligence and Applications (IJAIA), Vol.13, No.4, July 2022

using TensorFlowLite. Human supervision is not readily available in an industrial scenario to ensure the mask norms. This stand-alone system would help ensure that the mask mandates are followed with little human intervention. MdTobibul Islam et al.[54] developed an assistive technology for the visually impaired to identify the face of family members. They deployed a trained face-identification machine learning model to Raspberry Pi.

This work attempts to develop an efficient machine learning model for IoT devices by leveraging the recent developments in machine learning implementations on IoT devices and access to an elaborate open-access ECG dataset PTB-XL.

## 3. DATA PRE-PROCESSING

The PTB-XL dataset has 21837 records from 18885 patients. Each record is 10 seconds long and has 12 channels [55][56]. Meta-data such as sex, weight, height, and diagnostic class is also available. Statements about the signal category are present in 'scp_codes.' The dataset contains more information about the signal labels, which is out of the scope of this work. Although most signals are high quality [55], some signals occasionally contain powerline noise, baseline drift, burst noise, and static noise. Therefore, removing these artifacts is crucial before implementing the machine learning model.

### 3.1. Powerline Noise Removal

Occasionally, some channels present a powerline noise of 50-150 Hz. It is important to note that powerline noise may be present in some channels while absent in others for a given patient. For removing powerline noise, we have used a low pass filter with a cutoff at 45Hz with order 15, as the most significant frequencies for the machine learning model are under that. Using order 15th order ensures firm damping at the cutoff frequency.

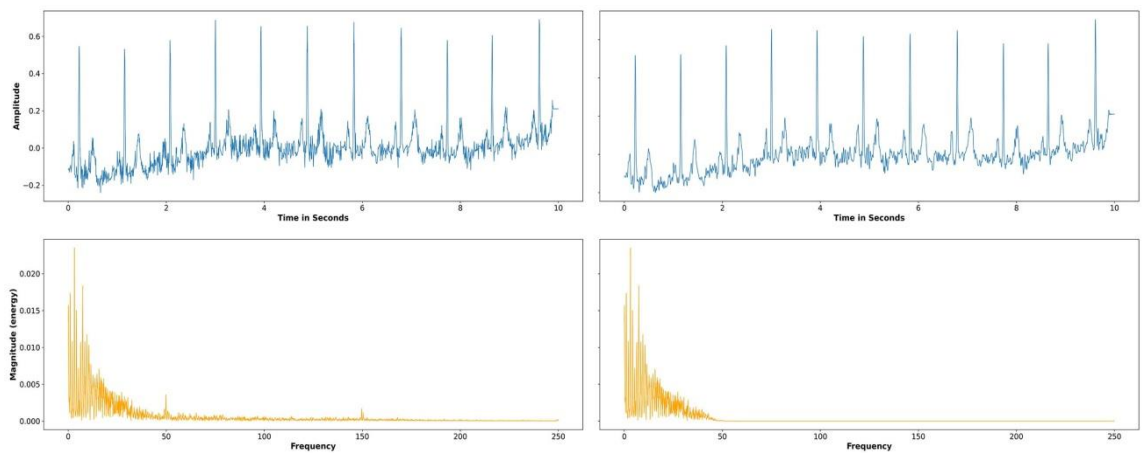

Figure 1. Spectrogram for ECG ID 1, Lead I - Unfiltered vs Filtered Signal

### 3.2. Baseline Wander Correction

A significant amount of baseline wander is present in many signals. Although some baseline wanders could be corrected using a high pass filter. This method causes a substantial reduction in the frequency domain, negatively impacting the machine learning model's performance. Therefore, we have used multiresolution decomposition based on wavelet transform. We determine the baseline for reconstructing the correcting signals. The determined baseline is

58

subtracted from the original signal to obtain the wander-free series. This method corrects the baseline while preserving important frequency information [57].

Fixed Signal = Original Signal – Baseline                                  *Eq. 1*

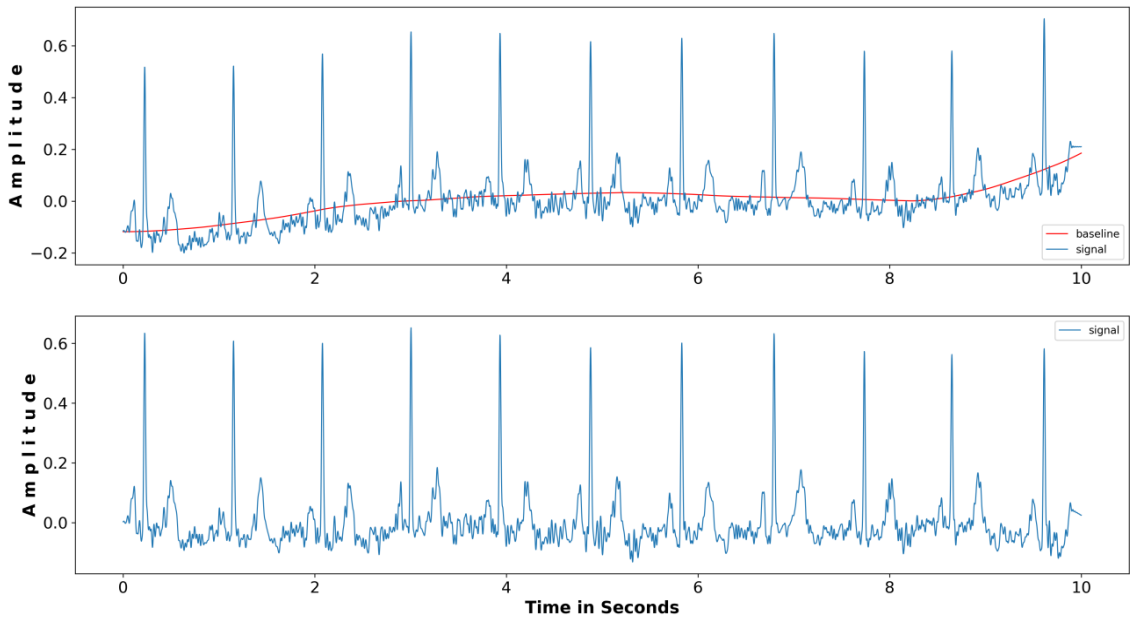

Figure 2. Baseline Wander Correction for ECG ID 1, Lead I

## 3.3. Rolling Mean

Rough peaks were present in some signals while absent in others belonging to the same classification. It would impact the model's learning performance. As a method of standardizing signals to some extent, the rolling mean was determined with a window size of 100 samples.

## 3.4. Determining Training Labels

The dataset lists SCP labels used in the scp_statements.csv file, which describes diagnostic class, form, and rhythm. The dataset mentions multiple SCP labels and their percentages of likelihood corresponding to each record. Although most of them are sorted and easy to interpret as they belong to a single superclass with high confidence, some are unsorted or have low confidence. ECG ID 39 and 63 are such example with labels – {'IMI': 15.0, 'LNGQT': 100.0, 'NST_': 100.0, 'DIG': 100.0, 'ABQRS': 0.0, 'SR': 0.0} and {'ASMI': 15.0, 'ABQRS': 0.0, 'SR': 0.0} respectively. This work selected labels with the highest confidence for the training process.

Fifty unique labels were identified and merged with their respective five superclasses. The superclasses as represented as Normal—NORM, ST/T change—STTC, Myocardial Infarction—CD, Hypertrophy—HYP, and Conduction Disturbance—MI. A new label 'OTHER' was created for signals not falling within the five superclasses. Following categorization was used to merge the labels-

NORM = ['NORM']
STTC = ['NDT', 'NST_', 'DIG', 'LNGQT', 'ISC_', 'ISCAL', 'ISCIN', 'ISCIL', 'ISCAS', 'ISCLA', 'ANEUR', 'EL', 'ISCAN' ]

MI = ['IMI', 'ASMI', 'ILMI', 'AMI', 'ALMI', 'INJAS', 'LMI', 'INJAL', 'IPLMI', 'IPMI', 'INJIN', 'INJLA', 'PMI', 'INJIL']
HYP = ['LVH', 'LAO/LAE', 'RVH', 'RAO/RAE', 'SEHYP']
CD = ['LAFB', 'IRBBB', '1AVB', 'IVCD', 'CRBBB', 'CLBBB', 'LPFB', 'WPW', 'ILBBB', '3AVB', '2AVB']
OTHER = ['AFLT', 'AFIB', 'PSVT', 'STACH', 'PVC', 'PACE', 'PAC']

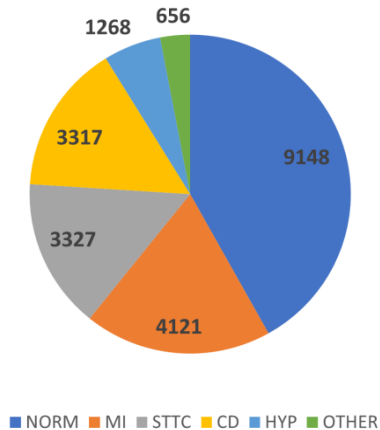
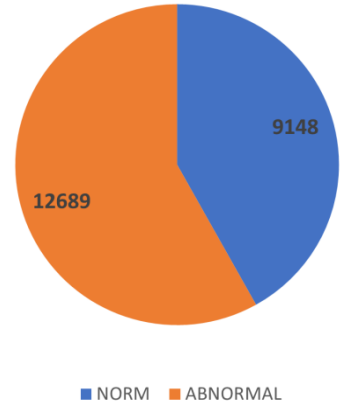

Figure 3. Distribution of occurrences after sorting and merging labels

Figure 4. Distribution of Normal and Abnormal Signals

The abnormal class was formed by merging MI, STTC, CD, HYP, and OTHER. 'NORM' created our 'normal' class. Although 'OTHER' is an imbalanced minority class, it is still used to estimate real-life implementation scenarios better. Since the data is imbalanced, class weights were used in the training process.

Class_weights = {0: 1.192181295358072, 1: 0.8611770524233432}

### 3.5. Training, Validation and Testing Splits

As suggested by the database authors, the data has been split into training, validation, and test sets [56]. Fold 1-8 forms the training data, fold nine forms the validation set, and fold ten to make the testing data. Since folds 9 and 10 were verified by a human, it would help determine the model's actual performance. Other methods that randomly split the data were not used. Some data had huge baseline wander. This baseline wander is beyond correction. Therefore we dropped those ECG ids.

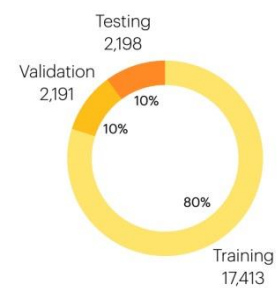

## 4. MODEL ARCHITECTURE

In this work, we have used multiple convolutional layers for extracting the features from the training data. We modeled with 1D CNN as they are more resource-efficient for our Lite Model. The model was designed with the view that it is lightweight and convertible to TensorFlowLiteFlatbuffer. Although some quantizations for this model are not well supported, we tried to balance between performance and resource constraints.

The input data is passed through 6 successive convolutional layers. The convolutional layers learn local spatial coherence, i.e., the local information present in 1D patches. This information is then used to identify similar patterns occurrences at other positions and signals. A Batch Normalization layer and a max-pooling layer with pool_size 2 follow each convolution layer. The variations in filter and kernel sizes help capture significant features from the input signal. The result of the convolutional layer is converted to a 1-D vector by passing them through a Flatten layer. Finally, the 1-D vector is processed by the two fully-connected layers. The final fully connected layer has one unit. The output of this dense layer provides the probability of a signal belonging to the class "Normal" or "Abnormal."

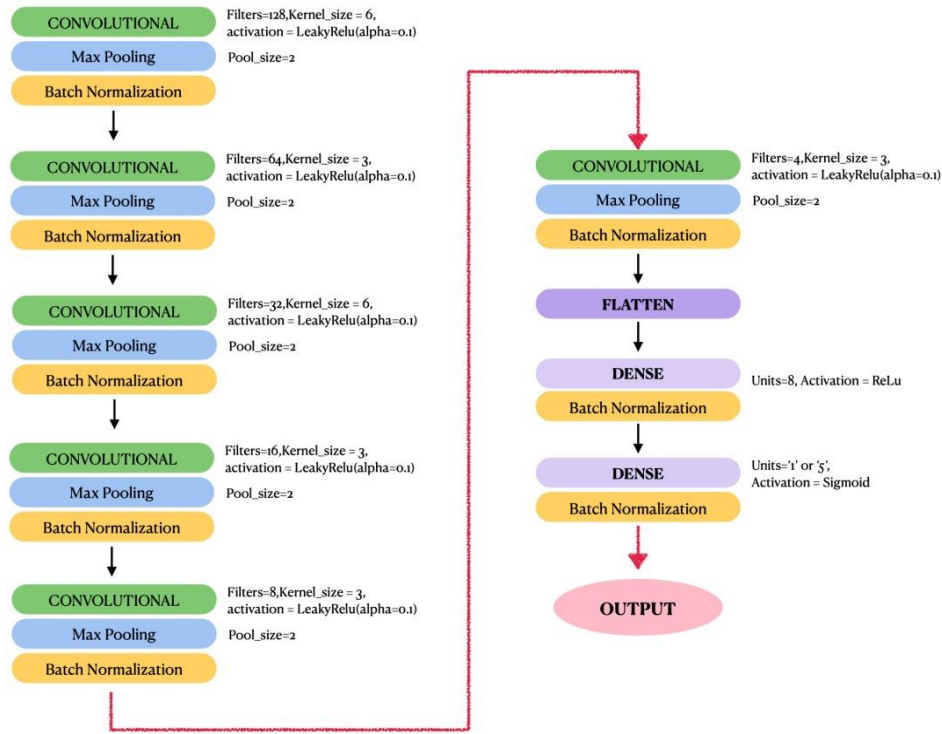

Figure 5. Model Architecture

Multiple batch-normalization layers were used to prevent initial random weight bias. Although we have tested numerous activation functions, 'LeakyReLu' yielded better results. Loss function 'binary_crossentropy' and 'adam' optimizer were used. Each model is run for 20 epochs. DataGenerator with batch size 32 was used for training on GPU. Inputs were shuffled with each batch to prevent the overfitting of data.

## 5. EVALUATION

We have trained the model using different leads to determine the best model for Raspberry Pi implementation. In our first model, we have used only lead I. We have used leads I, II, and III in our second model. In our third model, we have used leads I, II, III, AVL, AVR, and AVF. Finally, we have used all the leads for our fourth model.

Each model was saved and converted to TensorFlowlite. The TensorFlowLite model was run on Raspberry Pi 4.0 to evaluate its performance on the test dataset. TensorFlowLiteFlatBufferswere generated using all four trained models. Each FlatBuffer was quantized as 'float32' and 'float16'. Quantization optimizations beyond 'float16' are not easily possible due to 'LeakyReLu.'

Figure 8 and Table 1 show the performance of our model on the test dataset for all the models.

Table 2 presents the variations in size and accuracy of our models.

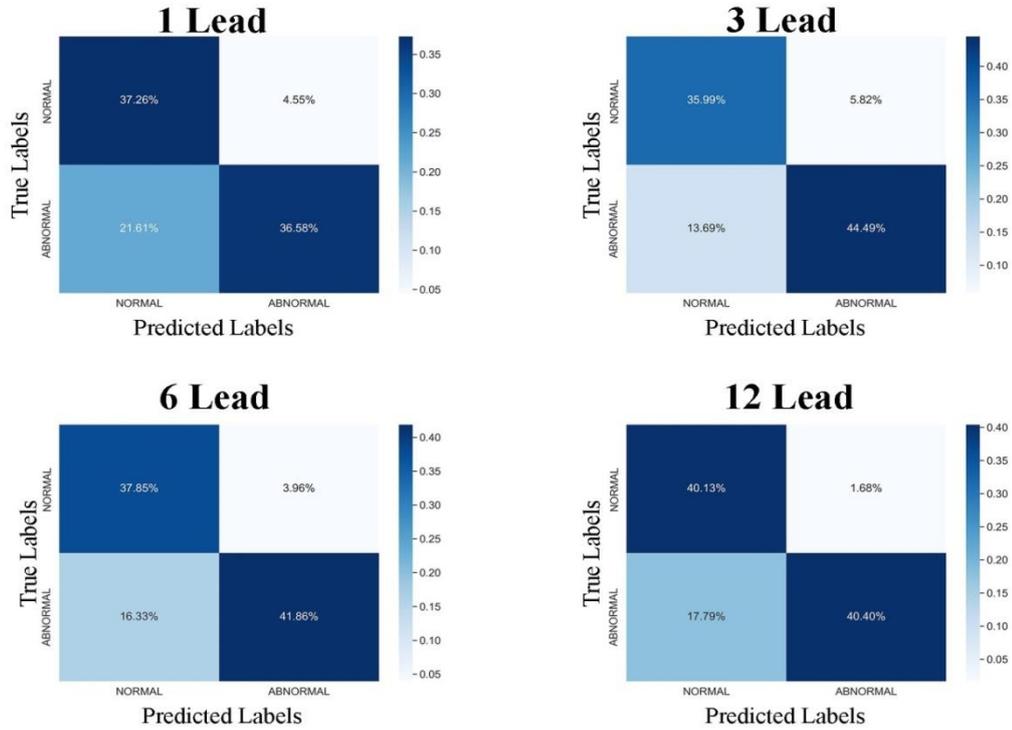

Figure 6. Confusion Matrix for all the Models

Table 1. TensorFlow/Keras Model Performance

| No. of Channels | Train Param | Accuracy % | Precision % | Recall % | F1 Score % |
|---|---|---|---|---|---|
| 1 | 60745 | 73.84 | 85.89 | 62.88 | 72.83 |
| 3 | 47901 | 80.48 | 88.36 | 76.21 | 81.37 |
| 6 | 49685 | 79.71 | 90.98 | 71.56 | 79.56 |
| 12 | 54293 | 81.21 | 93.75 | 71.94 | 81.21 |

Table 2. TensorFlowLite Performance on Raspberry Pi CM 4

| No. of Channels | TF Lite Float 32 | | TF Lite Float 16 | | Original Model |
|---|---|---|---|---|---|
| | Size (KB) | Accuracy (%) | Size (KB) | Accuracy (%) | Size (MB) |
| 1 | 197 | 73.84 | 112 | 73.83 | 1.38 |
| 3 | 203 | 80.48 | 115 | 80.52 | 1.40 |
| 6 | 212 | 79.70 | 120 | 79.70 | 1.44 |
| 12 | 230 | 81.21 | 128 | 81.21 | 1.50 |

## 6. CONCLUSIONS AND FUTURE WORK

One of the goals of this work was to develop an efficient deep learning model that could easily be deployed on resource-constrained IoT devices for ECG signal classification into 'Normal' and 'Abnormal' classes. Since PTB-XL is one of the largest open public databases available, our second goal was to evaluate its performance with our models.

This work demonstrated that using all twelve leads for classification yields the best accuracy. We can also observe that using at least three 'leads' over just one increases the classification accuracy by ~7%. However, precision increases when more than three 'leads' are used. But only a minor increase in accuracy is observed.

This work also demonstrates the use of TensorFlowLite for implementing machine learning models while maintaining accuracy. The accuracy of all the TensorFlowlite models was about the same as that of the original models. The 'float32' model presented a ~86% decrease in model size, whereas the 'float16' model presented a ~94% decrease.

Since TensorFlowlite may not support some features conventionally used in modeling, it is vital to explore more convertible models yielding higher accuracy. As an expansion of this work, we will be working on fused CNN-LSTM models. Additionally, we would focus on int8 and int16 quantizations for microcontroller implementations.

Source codes will be made available by the authors at [58].

## AUTHORS

**Kushagra Sharma** is a Graduate Research Assistant at the University of Manitoba. He joined Dr.RasitEskicioglu's lab as a graduate student in the Biomedical Engineering program to pursue his interest in the Internet of Things and Biomedical Instruments. His current research interests involve using machine learning to improve the productivity of low-powered mobile and embedded devices. He has also been exploring the signal analysis of gait motion, ECG, EEG, vestibular and cochlear nerves.

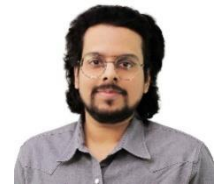

**Dr.RasitEskicioglu** is an Associate Professor at the computer science department, University of Manitoba, Canada. His research interests are primarily in experimental systems, particularly computer systems, systems software, operating systems, distributed, cluster, and grid computing, highspeed network interconnects, mobile networks, and pervasive computing. Currently, he is looking at wireless sensor networks (WSNs) and their applications for real-world problems, such as indoor localization, monitoring, and tracking, and the Internet of Things (IoT) for monitoring and tracking ageing populations.

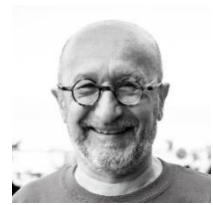